\documentclass[]{spie}  

\usepackage{amsmath,amsfonts,amssymb}
\usepackage{graphicx}
\usepackage[colorlinks=true, allcolors=blue]{hyperref}
\usepackage[T1]{fontenc}

\title{SPECULOOS: a network of robotic telescopes to hunt for terrestrial planets around the nearest ultracool dwarfs}

\author[a]{Laetitia Delrez}
\author[b]{Micha\"el Gillon}
\author[a]{Didier Queloz}
\author[c]{Brice-Olivier Demory}
\author[d,e]{Yaseen Almleaky}
\author[f]{Julien de Wit}
\author[b]{Emmanu\"el Jehin}
\author[g]{Amaury H. M. J. Triaud}
\author[h,b]{Khalid Barkaoui}
\author[b]{Artem Burdanov}
\author[i]{Adam J. Burgasser}
\author[b]{Elsa Ducrot}
\author[j,k]{James McCormac}
\author[a]{Catriona Murray}
\author[b,l]{Catarina Silva Fernandes}
\author[b]{Sandrine Sohy}
\author[a]{Samantha J. Thompson}
\author[b]{Val\'erie Van Grootel}
\author[m]{Roi Alonso}
\author[h]{Zouhair Benkhaldoun}
\author[m]{Rafael Rebolo}
\affil[a]{Cavendish Laboratory, JJ Thomson Avenue, Cambridge CB3 0HE, UK}
\affil[b]{Space Sciences, Technologies and Astrophysics Research (STAR) Institute, Universit\'e de Li\`ege, All\'ee du 6 Ao\^ut 19C, B\^at. B5C, 4000 Li\`ege, Belgium}
\affil[c]{University of Bern, Center for Space and Habitability, Sidlerstrasse 5, CH-3012 Bern, Switzerland}
\affil[d]{Space and Astronomy Department, Faculty of Science, King Abdulaziz University, 21589 Jeddah, Saudi Arabia}
\affil[e]{King Abdullah Centre for Crescent Observations and Astronomy, Makkah Clock, Mecca 24231, Saudi Arabia}
\affil[f]{Department of Earth, Atmospheric and Planetary Sciences, Massachusetts Institute of Technology, 77 Massachusetts Avenue, Cambridge, MA 02139, USA}
\affil[g]{School of Physics \& Astronomy, University of Birmingham, Edgbaston, Birmingham B15 2TT, UK}
\affil[h]{Laboratoire LPHEA, Oukaimeden Observatory, Cadi Ayyad University/FSSM, BP 2390 Marrakesh, Morocco}
\affil[i]{Center for Astrophysics and Space Science, University of California San Diego, La Jolla, CA 92093, USA}
\affil[j]{Dept. of Physics, University of Warwick, Gibbet Hill Road, Coventry CV4 7AL, UK}
\affil[k]{Centre for Exoplanets and Habitability, University of Warwick, Gibbet Hill Road, Coventry CV4 7AL, UK}
\affil[l]{University of Copenhagen, Centre for Star and Planet Formation, Niels Bohr Institute and Natural History Museum, DK-1350, Copenhagen, Denmark}
\affil[m]{Instituto de Astrof\'isica de Canarias, V\'ia L\'actea s/n, 38205 La Laguna, Tenerife, Spain, and Departamento de Astrof\'isica, Universidad de La Laguna, 38206 La Laguna, Tenerife, Spain}

\authorinfo{Further author information: (Send correspondence to L. Delrez and M. Gillon)\\L. Delrez: E-mail: lcd44@cam.ac.uk\\  M. Gillon (Project leader): E-mail: michael.gillon@uliege.be}

\begin{document} 
\maketitle

\begin{abstract}
We present here SPECULOOS, a new exoplanet transit search based on a network of 1m-class robotic telescopes targeting the $\sim$1200 ultracool (spectral type M7 and later) dwarfs bright enough in the infrared ($K$-mag $\leq 12.5$) to possibly enable the atmospheric characterization of temperate terrestrial planets with next-generation facilities like the {\it James Webb Space Telescope}.  The ultimate goals of the project are to reveal the frequency of temperate terrestrial planets around the lowest-mass stars and brown dwarfs, to probe the diversity of their bulk compositions, atmospheres and surface conditions, and to assess their potential habitability. 

\end{abstract}


\section{Introduction}
\label{sec:intro}  

One of the most fundamental questions raised by humankind is how frequently, and under which conditions life exists around other stars. Its scientific answer requires the detailed atmospheric characterization of temperate rocky exoplanets to search for potential biosignature gases\cite{seager2016}. The sample of transiting exoplanets found so far represents a genuine Rosetta stone to understand extrasolar worlds because their special geometry offers the possibility to study their atmospheres through eclipse (transit and occultation) spectroscopy, without having to spatially resolve them from their host stars \cite{deming2017}. Over the last $\sim$15 years, these techniques have been applied to transiting hot Jupiters, Neptunes, and a few favorable Super-Earths, giving us first insights into their atmospheric chemical compositions, pressure-temperature profiles, albedos, and circulation patterns (for comprehensive reviews, see e.g. Ref. \citenum{deming2017} or \citenum{crossfield2015}). However, exporting these methods to the atmospheric characterization of an Earth-twin transiting a Sun-like star is out of reach for all the existing or currently planned astronomical facilities. The main reasons for this are the lack of a suitable target, the overwhelmingly large area contrasts between the solar disk and the tiny Earth's atmospheric annulus, and between the Sun's and Earth's luminosities. They lead to very-low signal-to-noise ratios (SNRs) for any atmospheric spectroscopic signature and any realistic observing program with upcoming facilities such as the \textit{James Webb Space Telescope}\cite{gardner2006} (JWST)\cite{Kaltenegger2009}. Fortunately, this negative conclusion does not hold for an Earth-sized planet transiting a nearby host of the smallest kind - a nearby `ultra-cool dwarf'\cite{kirkpatrick2005} (UCD).

In this context, we present here our ground-based transit survey SPECULOOS\footnote{\url{http://www.speculoos.uliege.be}} (Search for Planets EClipsing ULtra-cOOl Stars, Burdanov et al. 2017\cite{burdanov2017}, Gillon et al. 2018\cite{gillon2018}), which aims at exploring the nearest UCDs for transits of temperate terrestrial planets, focusing on those that are bright enough in the near-IR to make possible the atmospheric study of a transiting Earth-sized planet with JWST and other future facilities.

\section{SPECULOOS: seizing the ultracool dwarfs opportunity}
\label{sec:speculoos}

UCDs are traditionally defined as dwarfs with effective temperatures lower than 2700 K, luminosities lower than 0.1\% that of the Sun, and spectral types M7 and later, including L, T and Y dwarfs\cite{kirkpatrick2005,cushing2011}. They have masses below $\sim$0.1 $M_{\odot}$, extending below the hydrogen burning minimum mass (HBMM) of 0.07 $M_{\odot}$, into the realm of non-fusing brown dwarfs. Their sizes are comparable to that of Jupiter ($\sim$0.11 $R_{\odot}$), with radii reaching a minimum of $\sim$0.08-0.10 $R_{\odot}$ near the HBMM. Their cool atmospheres are rich in molecular gases (H$_2$O, CO, TiO, VO, CH$_4$, NH$_3$, CaH, FeH) and condensed refractory species (mineral and metal condensates, salts and ices), producing complex spectral energy distributions, strongly influenced by composition and chemistry, that peak at near- and mid-IR wavelengths. Thanks to the proliferation of red optical and near-IR surveys over the past 20 years, the UCDs population in the close solar neighbourhood (<20 pc) is now well known\cite{reid2007,reid2008}, except for the coolest (late-T and Y types) brown dwarfs. Overall, UCDs appear to be less numerous than more massive M stars (0.1 $M_{\odot}$ < mass < 0.5 $M_{\odot}$), but are more abundant than solar-type FGK stars in the immediate solar neighbourhood (e.g., >5:1 UCDs:G dwarfs in the 8 pc census)\cite{kirkpatrick2012}.

Despite that UCDs represent a significant fraction of the Galactic population, their planetary population is still a nearly uncharted territory. As of today, only nine \textit{bona fide} planets have been found in orbit around UCDs: two planets detected by microlensing around distant UCD stars -- the $\sim$3.2 $M_{\oplus}$ planet MOA-2007-BLG-192Lb\cite{kubas2012}, and the $\sim$1.4 $M_{\oplus}$ planet OGLE-2016-BLG-1195Lb\cite{shvartzvald2017} -- and the seven Earth-sized planets transiting the nearby UCD star TRAPPIST-1\cite{gillon2016,gillon2017} (see Section \ref{sec:t1}). The two microlensing planets are very interesting, because they demonstrate that UCDs can form planets more massive than the Earth despite the low mass of their protoplanetary discs. On their side, the Earth-sized planets transiting TRAPPIST-1 indicate that compact systems of small terrestrial planets are probably common around UCDs, as TRAPPIST-1 is one of only $\sim$50 UCDs targeted by the SPECULOOS prototype survey ongoing on the TRAPPIST-South telescope\cite{gillon2011,jehin2011} since 2011 (see Section \ref{sec:prototype}). Objects of one to a few Jupiter masses have also been found around UCDs by microlensing\cite{han2013,han2016,jung2018} and direct imaging \cite{chauvin2004,todorov2010,gauza2015,liu2011}, but they were probably formed similarly to binary systems through gravitational fragmentation\cite{lodato2005}, and should thus be considered more as ``sub-brown dwarf'' than as planets\footnote{The case of the three microlensing giant planets\cite{han2013,han2016,jung2018} is not crystal clear, because their orbits are tight enough (0.9, 0.7, and 0.6 AU, respectively) to leave possible a formation within a protoplanetary disk. Still, their high masses ($\sim$2, $\sim$0.7, and $\sim$0.7 $M_{\mathrm{Jup}}$, respectively) make such a scenario unlikely, given the small masses of UCD disks.}. 

The planet detections mentioned above are consistent with the growing observational evidence that young UCDs are commonly surrounded by long-lived protoplanetary discs\cite{scholz2008,luhman2012} and also exhibit the hallmarks of pre-planetary formation: evidence of disc accretion\cite{muzerolle2000}, circumstellar dust excess\cite{klein2003}, accretion jets\cite{whelan2005}, and planetesimal formation\cite{pascucci2011,ricci2013}. Furthermore, while they do not favor the formation of giant planets around UCDs, planetary formation models do predict terrestrial planets\cite{payne2007,raymond2007,montgomery2009}, with possible volatile-rich compositions depending on the structure and evolution of the planetary disc\cite{alibert2017}. These observational and theoretical considerations hint at the presence of a large, mostly untapped, population of low-mass planets around UCDs.

Radial velocity and transit surveys have revealed that close-in
packed systems of low-mass planets are very frequent around solar-type
stars\cite{mayor2011,fressin2013}. Half of the stellar content in our Galaxy is composed of stars with a mass inferior to 0.25 $M_\odot$ \cite{henry06}, which is approximately three times more than single Sun-like stars. Early statistics also indicate that rocky planets might occur three to five times more frequently around such stars than Solar twins\cite{bonfils2013,dressing2015,he2017}. If this trend continues to the bottom of the main-sequence, low-mass planets around UCDs could well represent the most common Earth-sized planets in our Galaxy. The detection, and the study of these planets is therefore crucial to comprehend how Earth-like planets assemble, which range of climatic conditions are possible, and which variety of atmospheric compositions exist. Regardless of whether those worlds turn out to be habitable, or indeed inhabited, their detection and their study will inform us about the frequency and conditions for biology in the Cosmos.

\begin{figure} [ht]
\begin{center}
\includegraphics[height=5.28cm]{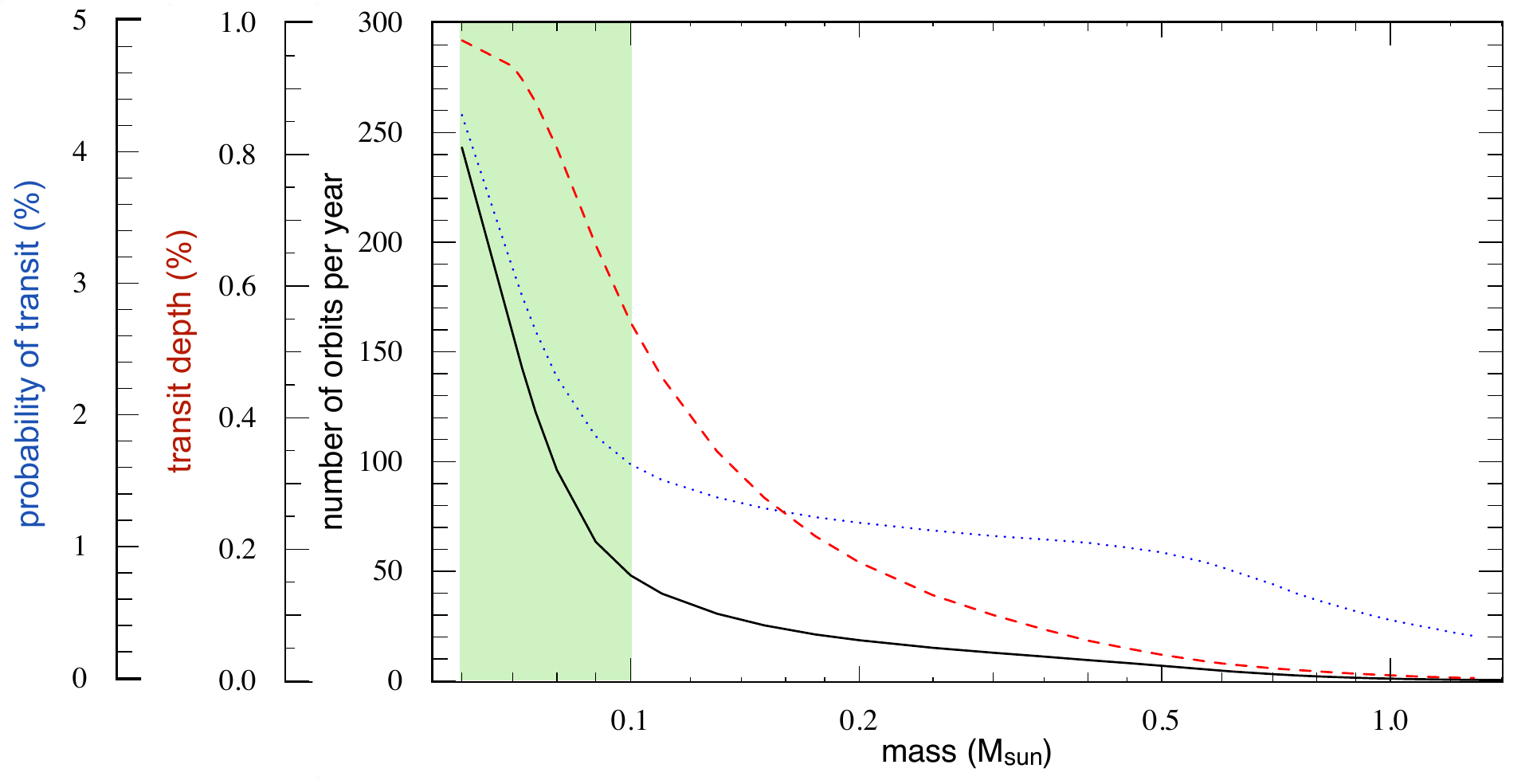}
\includegraphics[trim=0 -0.3cm 0 0, height=5.27cm]{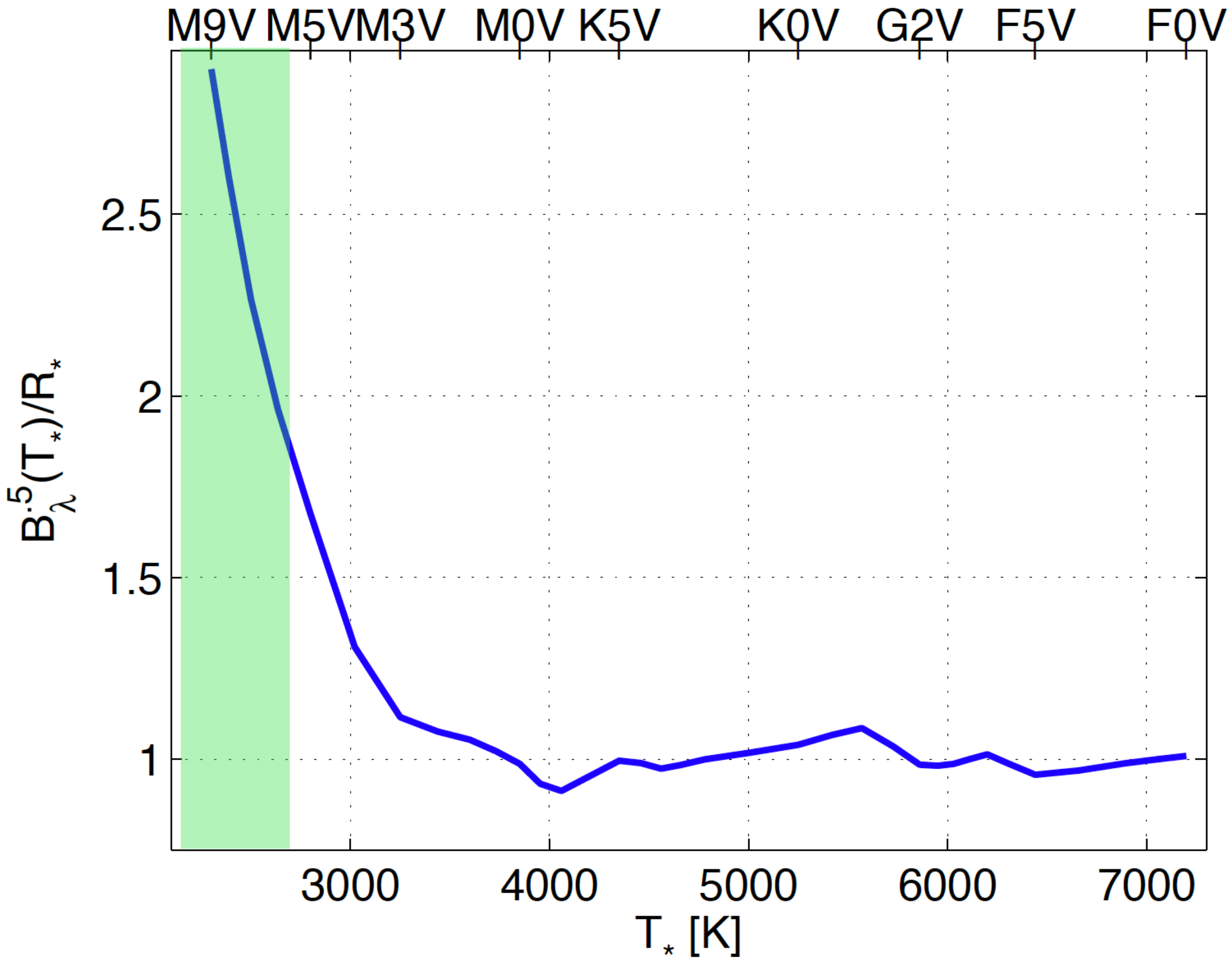}
\end{center}
\vspace{-0.3cm}
\caption{\textit{Left:} Probability of transit (dotted blue), transit depth (dashed red), and number of orbits (i.e. number of transits or occultations) per year (black) of an Earth-sized planet with an equilibrium temperature of 255 K (like Earth) as a function of the host's mass. Figure adapted from He et al. (2017)\cite{he2017,triaud2013}. \textit{Right:} Ratio $\sqrt{B_{\lambda}(T_{\star})}/(R_{\star})$ -- normalized for a Sun-like star -- as a function of the effective temperature $T_{\star}$. For a given planet (fixed radius, mass, and equilibrium temperature), the SNR of an atmospheric spectral feature in transmission scales as this ratio. Figure adapted from de Wit \& Seager (2013)\cite{dewit2013}. For both panels, the green area shows approximately the range that SPECULOOS will explore.}
\label{fig:ucd}
\end{figure}

UCDs have characteristics that make them not only ideal targets to
search for transiting temperate rocky worlds, but also optimal for their atmospheric characterization. Figure \ref{fig:ucd} (left) shows the probability of transit, transit depth, and number of orbits per year for an Earth-sized planet, with an equilibrium temperature of \hbox{255 K} (like Earth), as a function of the host's mass. It can be seen that UCDs present several advantages. First, their small size leads for Earth-sized planets to transit signals that are about two orders of magnitude deeper than for Earth-Sun twin systems. Expected transit depths range from a few 0.1\% up to >1\%, which is routinely detected by many ground-based surveys\cite{pollacco2006}. Because of their low effective temperature, UCDs also harbor a habitable zone\cite{kopparapu2013} that is much closer to them than that of a solar-type star, making the transits of potentially habitable planets more likely (geometric transit probabilities higher than 1.5\%) and frequent (periodicities of a few days, $\gtrsim$100 transits/year). UCDs are also optimal targets for the atmospheric characterization of their potential transiting planets, thanks to their low luminosity and small size, and the resulting large planet-to-star flux and size ratios. Regarding for example the transmission spectroscopy technique, de Wit \& Seager (2013)\cite{dewit2013} showed that for a given planet (fixed radius, mass, and equilibrium temperature), the SNR of an atmospheric spectral feature scales as $\sqrt{B_{\lambda}(T_{\star})}/(R_{\star}d)$, where $B_{\lambda}(T_{\star})$ is the host's spectral radiance (with $B_{\lambda}$ the Planck function and $T_{\star}$ the effective temperature), $R_{\star}$ is the host's radius, and $d$ is its distance to Earth. Figure \ref{fig:ucd} (right) shows the ratio $\sqrt{B_{\lambda}(T_{\star})}/(R_{\star})$ -- normalized for a Sun-like star -- as a function of the effective temperature $T_{\star}$. This ratio -- and thus the SNR in transmission -- is constant for all stars with earlier spectral types than M2V, but increases significantly towards late M dwarfs and peaks for UCDs. As a consequence, fewer transit observations need to be co-added to reach a significant detection of atmospheric features for terrestrial planets transiting UCDs than for those around solar-type stars. In this regard, the fast occurrence of eclipses mentioned above for potentially habitable planets around UCDs is also a considerable advantage, as it takes much less time to obtain a given amount of in-transit observations for these planets than for an Earth-Sun twin system. The SNR scales inversely as the distance to the host, making the nearest UCDs the best targets for the search of temperate rocky exoplanets on which potential biosignatures could be detected with the next-generation facilities. This is the main goal of our SPECULOOS survey.

\begin{figure} [ht]
\begin{center}
\includegraphics[height=5.9cm]{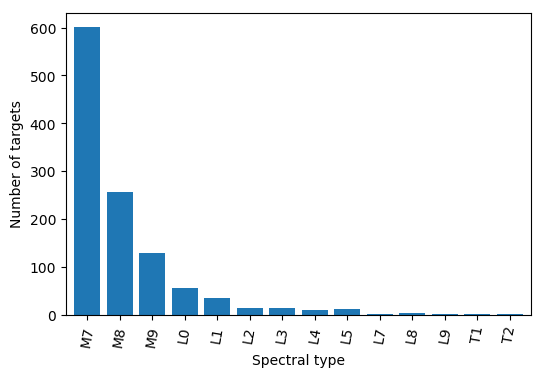}
\includegraphics[trim=0 +1.63cm 0 0, height=5.9cm]{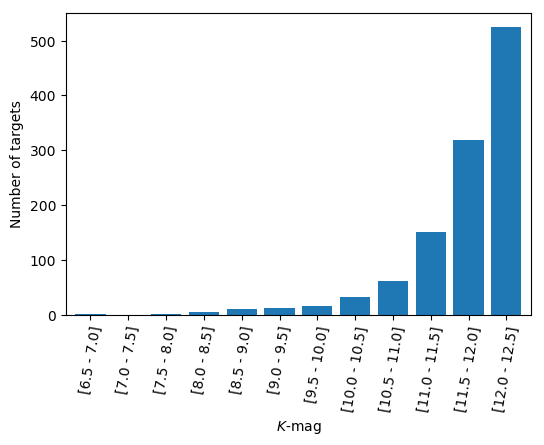}
\end{center}
\vspace{0.6cm}
\caption{Histograms of the SPECULOOS targets as a function of their spectral type (\textit{left}) and $K$-band magnitude (\textit{right}).}
\label{fig:targets}
\end{figure} 

\section{Survey design} 

\subsection{Target sample}

The target sample of our survey consists of all the known nearby UCDs with a $K$-band magnitude $\leq$ 12.5, i.e. bright enough in the near-IR for possible future follow-up studies with JWST. Following de Wit \& Seager (2013)\cite{dewit2013}, an Earth-sized planet around an M9 dwarf star is amenable to atmospheric characterization via transmission spectroscopy up to 50 pc assuming 200 hours of in-transit observations. Assuming for SPECULOOS' scope a cap for JWST programs targeting temperate rocky planets of 150 hours (and noting that TRAPPIST-1's $K$-mag = 10.3), we fixed our targets' brightness limit to $K$-mag = 12.5. We identified $\sim$1200 possible targets, among which $\sim$86.7\% M-type (M7-M9), $\sim$13.2\% L-type, and $\sim$0.02\% T-type dwarfs. Figure \ref{fig:targets} shows their distribution as a function of spectral type and $K$-mag. The vast majority of our targets are UCD stars. Based on their spectral type distribution, we indeed estimate that only $\sim$10\% of our targets are brown dwarfs.

\subsection{General instrumental concept and observational strategy}
\label{sec:concept}

Of course, these UCD targets are spread all over the sky, which means that they have to be monitored individually. However, the short orbital periods of planets in the habitable zone around UCDs (cf. Section \ref{sec:speculoos}) translate into a required photometric monitoring of much smaller duration for the detection of such planets than for Earth-Sun twin systems. Consequently, a transit search targeting the $\sim$1200 UCDs of our sample could be done within $\sim$10 years with a realistically small number of ground-based telescopes. Still, this strategy makes necessary being able to detect a single transit event to prevent allocating one telescope to one UCD for unrealistic durations, and thus results in a stringent requirement for very-high photometric precisions ($\lesssim$0.1\%).

Due to their low temperatures, UCDs are faint in the optical, their spectral energy distribution peaking at near- and mid-IR wavelengths. Using large telescopes combined with infrared detectors could thus appear \textit{a priori} as the only option for achieving high photometric precisions on these objects. SNR computations convinced us that it was not the case, and that 1m-class telescopes equipped with near-IR-optimized CCD cameras (providing good efficiencies up to 1 $\mu$m) should reach the required high photometric precisions. We validated this strategy through a prototype survey, that we will describe in Section \ref{sec:prototype}.

Due to our requirement for very-high photometric precisions, we choose to observe a given target continuously during a number of nights long enough to explore efficiently its habitable zone for transiting planets, instead of making each telescope point another target every few minutes in the course of the nights, as done by some other surveys\cite{Nutzman2008}. The continuous observation of the targets does not only maximize the photon counts, it also minimizes systematics and improves the photometric detection threshold by letting the telescope keep the stars on the same pixels of the detector during the whole night, thus optimizing the capacity to detect low-amplitude transits. Furthermore, the need for continuous observation is also driven by the expected short transit duration (down to 15 min) for very-short-period ($\leq$1 d) planets orbiting UCDs.

Each target should be observed during a number of nights long enough to explore efficiently its habitable zone for transiting planets. For our survey, we choose to fine-tune this monitoring duration for each target as a function of its spectral type so as to reach a 70\% probability of observing the transit of a planet that receives the same irradiation from its host star as the Earth does from the Sun. This was done by simulations (see Section 7.1) that resulted in monitoring durations per target of 20, 16, 13, and 10 nights for the spectral types M7, M8, M9, and later, respectively (assuming 8 hours of continuous observation per night). Integrating over our $\sim$1200 targets results in a total of 19,500 nights required by the survey, what can be done in $\sim$10 years considering realistic observations with a network composed of two facilities of 4 telescopes each, one in each hemisphere, and assuming a global efficiency of 70\% (i.e. 30\% of loss due to, e.g., bad weather or technical problems).

These considerations sketch the general instrumental concept of our survey: a network of ground-based 1m-class optical telescopes equipped with near-IR-optimized CCD cameras, monitoring each UCD of our sample individually and continuously, for a duration long enough to probe efficiently its habitable zone for transiting planets.

\subsection{A global network of robotic telescopes}

SPECULOOS will eventually be based on two nodes, one in each hemisphere. The southern one, the SPECULOOS Southern Observatory (SSO), is currently being commissioned at the ESO Paranal Observatory in the Chilean Atacama Desert. It consists of four 1-m telescopes that will explore our $\sim$600 southern UCD targets for transits. The northern node of SPECULOOS, the SPECULOOS Northern Observatory (SNO), is also planned to consist of four 1-m telescopes, and will be located at Teide Observatory in Tenerife (Canary Islands). SAINT-EX\footnote{\url{http://www.saintex.unibe.ch/}}, a new robotic 1-m telescope being installed at the National Astronomical Observatory of Mexico (San Pedro M\'artir), will also partially contribute to SPECULOOS. Finally, our two 60-cm robotic telescopes, TRAPPIST-South (ESO La Silla Observatory, Chile) and TRAPPIST-North (Ouka\"imeden Observatory, Morocco)\footnote{\url{http://www.trappist.uliege.be/}}, also participate in the SPECULOOS survey, focusing on its $\sim$100 brightest targets.

\section{Prototype on TRAPPIST-South}
\label{sec:prototype}

SPECULOOS started back in 2011 as a prototype mini-survey\cite{gillon2013ucdts} on our TRAPPIST-South telescope\cite{jehin2011,gillon2011}, a 60-cm Ritchey-Chretien telescope that has been installed since 2010 at ESO La Silla Observatory in the Chilean Atacama Desert. The telescope is equipped with a 2K$\times$2K thermoelectrically-cooled back-illuminated CCD camera offering excellent quantum efficiencies from 300 to beyond 900 nm. It has a field of view of 22'$\times$22' and a pixel scale of 0.64''/pixel. This prototype survey targets 50 among the brightest southern UCDs, with $K$-magnitude between 5.3 and 11.4 (mean $K$-mag=9.7), and uniformly distributed in terms of spectral type and sky position. Its concept is to monitor in a wide near-IR filter (transmittance > 90\% from 750 nm) each UCD during at least 100 hours spread over several nights. The initial goals of this prototype survey were to assess the typical photometric precisions that can be reached for UCDs on nightly timescales with such an instrumental setup and the resulting detection thresholds for terrestrial planets.

Importantly, we also wanted to identify the possible astrophysical and atmospheric limitations of the SPECULOOS concept. Indeed, UCDs are commonly considered as variable objects\cite{goldman2005,reid2005}. This is related to stellar activity for UCD stars (flaring emission, rotational modulation induced by spots on the photosphere) and to evolving weather patterns for brown dwarfs (rotational modulation due to heterogeneous cloud covers). This variability could have been a big issue for our survey, strongly limiting the ability to detect low-amplitude transits\cite{reid2005}. Another possible barrier to the relevance of the SPECULOOS concept could have come from the Earth's atmosphere. Indeed, in the very-near-IR, the water molecule and OH radical contribute to a number of absorption bands, as well as significant emission for OH (airglow). This brings unavoidable significant levels of red noise in photometric time-series\cite{berta2011}, whose impact on our ability to detect transits had to be thoroughly assessed.

\subsection{Results}

\begin{figure} [ht]
\begin{center}
\includegraphics[height=11.9cm]{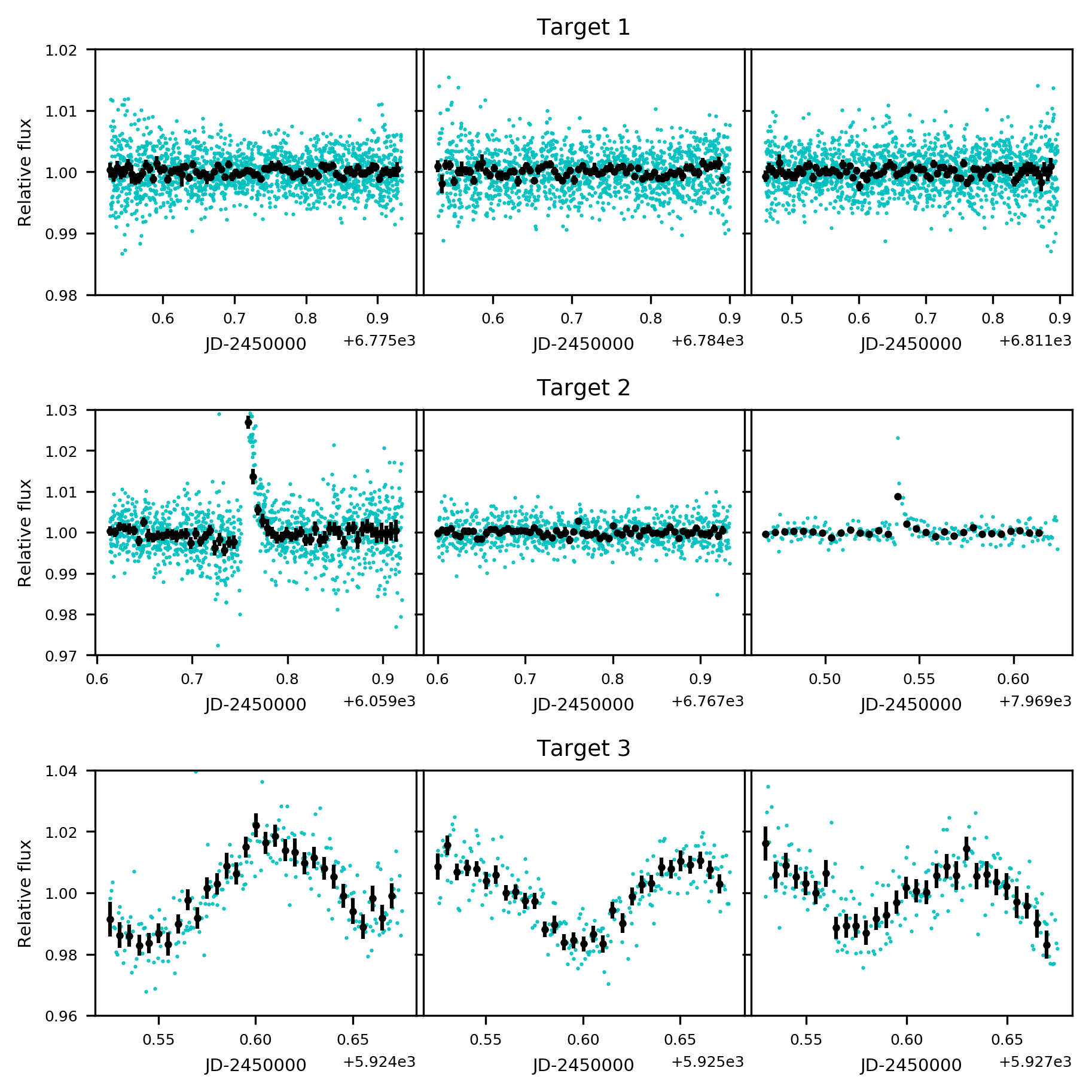}
\end{center}
\vspace{-0.2cm}
\caption{Typical TRAPPIST-South light curves of UCDs obtained as part of our prototype survey. We consider 3 different UCDs and present for each of them 3 light curves, obtained over different nights. For each light curve, the measurements are shown unbinned (cyan points) and binned per 7.2 min (black points with error bars). The first target (\textit{top}) shows rather flat light curves. The second one (\textit{middle}) shows clear flares in some light curves (see the first and third light curves shown here). The last target (\textit{bottom}) shows some rotational modulation with an amplitude of $\sim$4\% and a period of $\sim$2.8 hours.}
\label{fig:ucdts}
\end{figure}

About 40 UCDs were observed by TRAPPIST-South in the period from 2011
to 2017. Half of the observed UCDs show ``flat'' light curves, i.e. stable photometry on the night timescale (see Fig. \ref{fig:ucdts} - top). Some of the other UCDs ($\sim$20\%) show clear flares in some light curves. These flares are seen in near-IR light curves as sudden increase of a few percents of the measured brightness, followed by a gradual decrease back or close to the normal level, the whole process taking only 10 to 30 min (see Fig. \ref{fig:ucdts} - middle). In the context of a transit search, it is easy to identify and discard the portions of light curves affected by flares. Furthermore, their frequency is relatively small (1 flare per 3-4 nights on average). Finally, about 30\% of the observed UCDs show some rotational modulation (and more complex variability) with up to 5\% amplitude (see Fig. \ref{fig:ucdts} - bottom). These rotational modulations do not limit the ability to detect transits as they can be modeled and corrected. One of these objects is the nearby brown dwarf binary Luhman 16AB, that we monitored for nearly a fortnight as part of our prototype survey, right after its discovery was announced in February 2013\cite{luhman2013}. The quality of our photometric data allowed us to reveal fast-evolving weather patterns in the atmosphere of the coolest component of the binary, as well as to firmly discard the transit of a two-Earth-radii planet over the duration of the observations and of an Earth-sized planet on orbits shorter than $\sim$9.5 hours\cite{gillon2013luhman}.

From simulations based on the injection and recovery of synthetic transits of terrestrial planets in TRAPPIST-South UCD light curves, we have reached the conclusion that the variability of a fraction of UCDs (flares and rotational modulation) does not limit the ability to firmly detect transits of close-in Earth-sized planets. The achieved photometric precisions are globally nominal, with no hint of extra correlated noise, except when the observations were performed in high-humidity conditions. Near-IR ground-based CCD time-series photometry of UCDs from a suitable astronomical site (good transparency, low humidity) can thus reach nominal sub-mmag precisions. These conclusions were recently strengthened by our detection of an amazing planetary system around one of the TRAPPIST-South UCD target, TRAPPIST-1 (Gillon et al. 2016\cite{gillon2016}, 2017\cite{gillon2017}).

\label{sec:t1}
 
The nearby ($\sim$12 pc) TRAPPIST-1 system is composed of a middle-aged\cite{burgasser2017} M8-type dwarf star\cite{vangrootel2018} orbited by seven nearly Earth-sized\cite{delrez2018} planets in orbits of 1.5 to 19 days\cite{gillon2016,gillon2017,luger2017}. All seven planets are temperate (equilibrium temperatures <400 K assuming a null Bond albedo) and the three planets TRAPPIST-1e, f, and g orbit in the habitable zone around the star (as estimated following Ref. \citenum{kopparapu2013}). Remarkably enough, the seven planets form a (near-)resonant chain, with all sets of three adjacent planets in Laplace three-body resonances\cite{luger2017}. This particular configuration greatly enhances the gravitational interactions between the planets and the resulting transit timing variations (TTVs, see e.g. Ref. \citenum{agol2005} or \citenum{holman2005}), whose dynamical modeling makes it possible to constrain the bulk densities and eccentricities of the planets\cite{grimm2018}. The TRAPPIST-1 planets have thus become prime targets for the study of temperate terrestrial worlds outside the Solar System, including the study of their atmospheres, owing to their transiting nature, combined with the infrared brightness ($K$-mag=10.3), Jupiter-like size ($\sim$0.12 $R_{\odot}$), and low luminosity ($\sim$0.0005 $L_{\odot}$) of their host star\cite{barstow2016,dewit2016,morley2017,dewit2018}. Since the publication in February 2017 of our discovery paper announcing seven temperate terrestrial planets around TRAPPIST-1 (Gillon et al. 2017\cite{gillon2017}), more than 200 studies have approached this unique system. An interesting aspect is the sheer diversity across the scientific disciplines that are interested in the TRAPPIST-1 system (planet formation and evolution, astrobiology, geophysics, atmospheric science, ...). This demonstrates the strong scientific potential of our SPECULOOS project, with expected results that will benefit a broad community in a field that is becoming increasingly multidisciplinary.

\setcounter{footnote}{0}

\begin{figure*} [ht]
\begin{center}
\includegraphics[height=6.0cm]{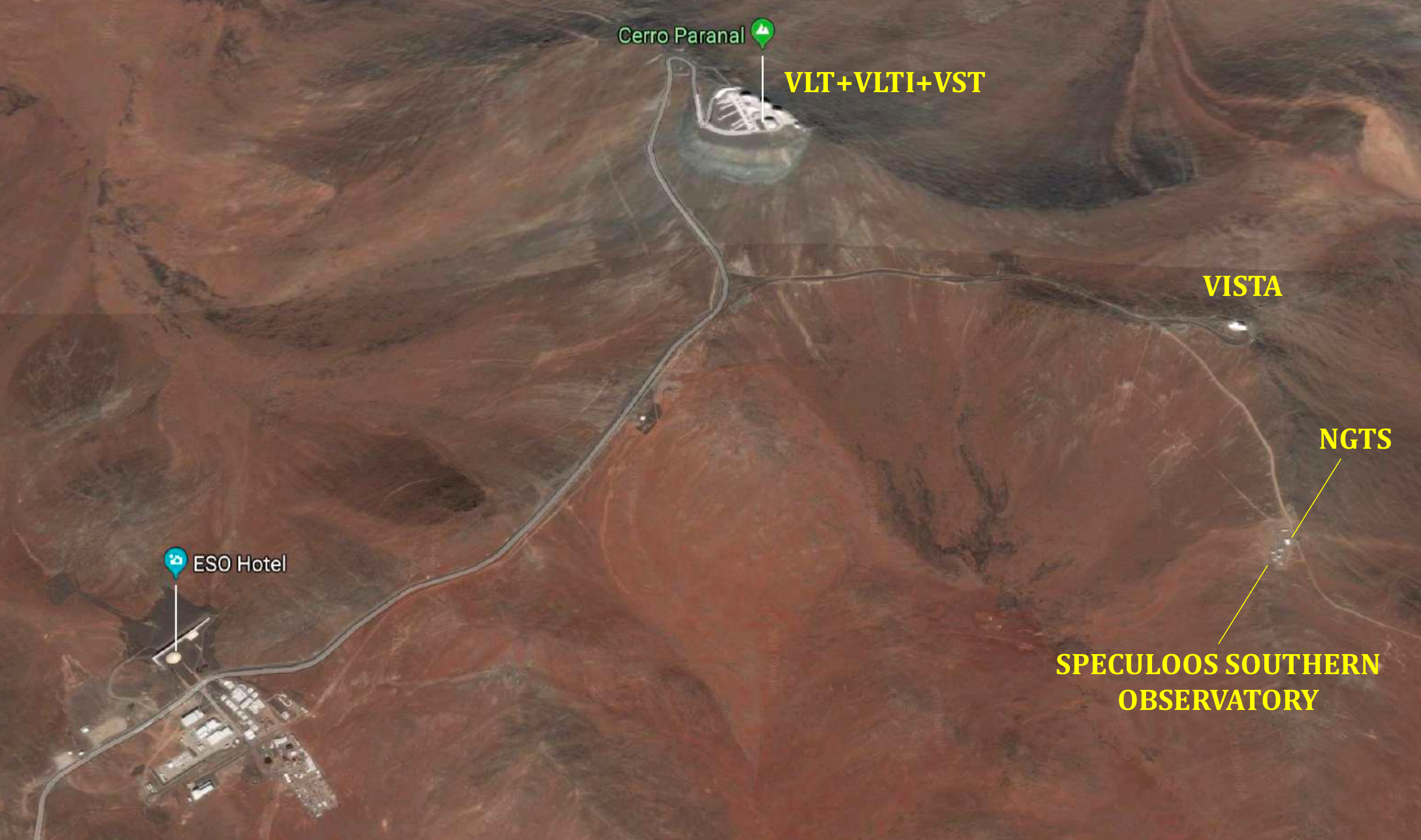}
\includegraphics[height=7.5cm]{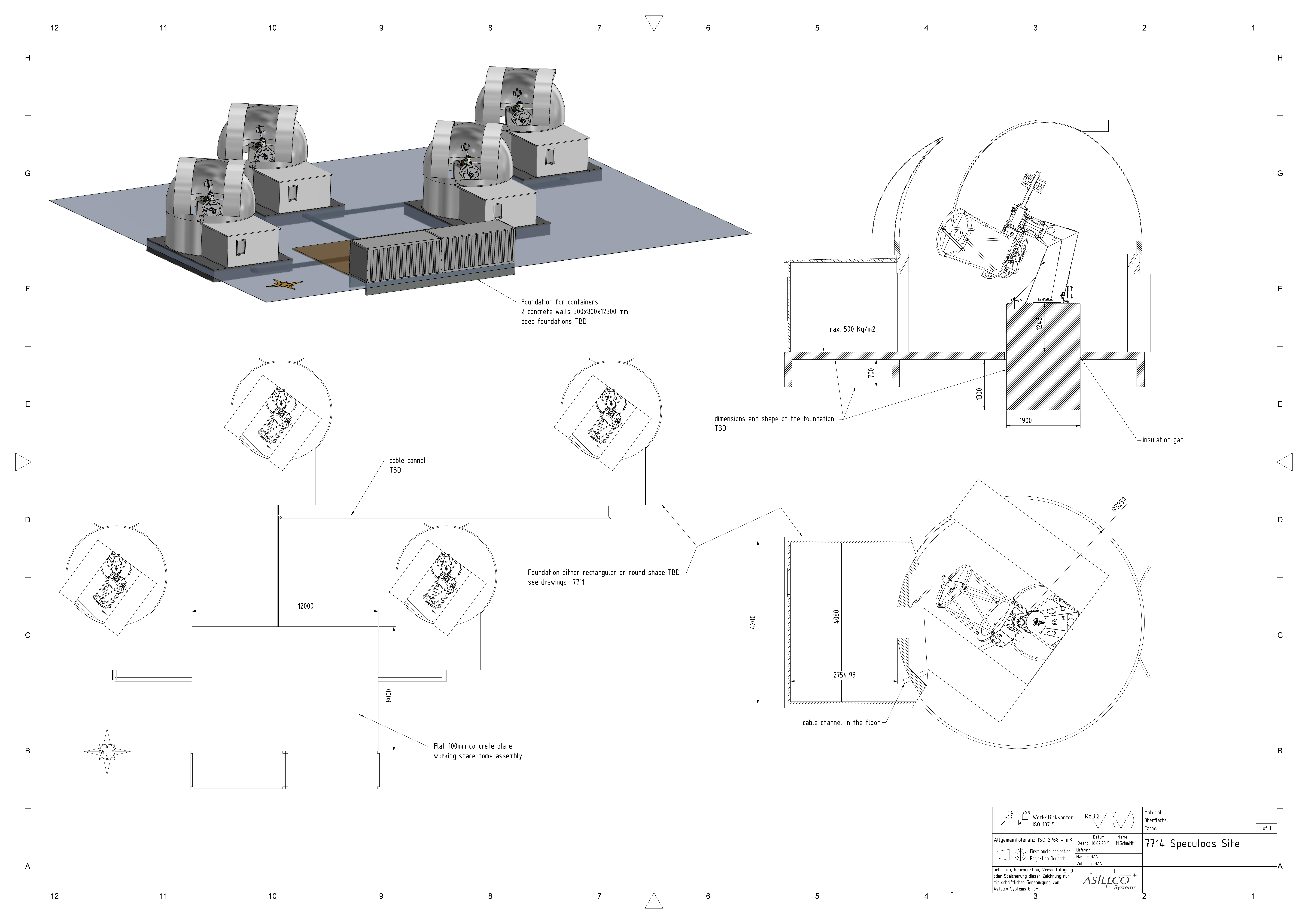}
\end{center}
\vspace{-0.2cm}
\caption{Location of the SPECULOOS Southern Observatory at ESO Paranal Observatory (\textit{top}, credit: Google Maps) and technical drawing of the facility (\textit{bottom}, credit: ASTELCO).}
\label{fig:sketch}
\end{figure*} 


\begin{figure*} [!ht] 
\vspace{0.5cm}
\begin{center}
\includegraphics[height=5.4cm]{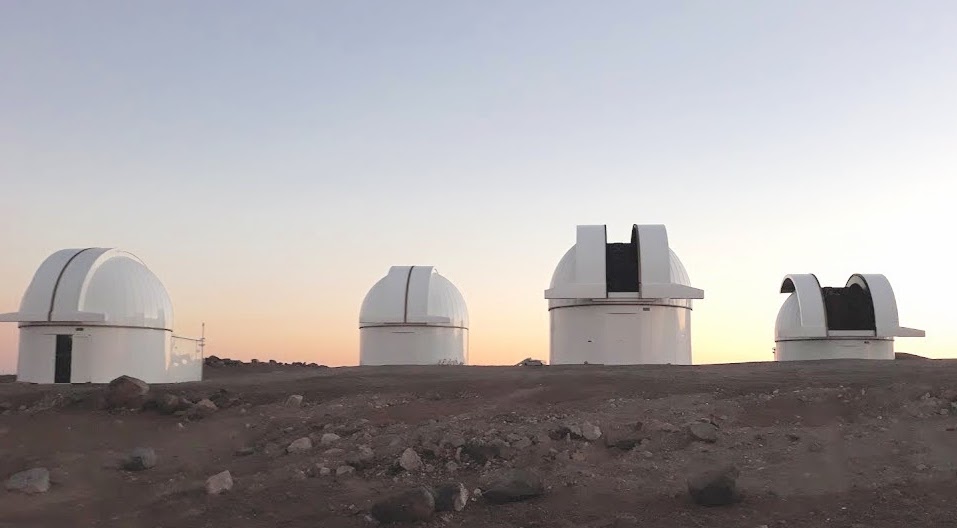}
\includegraphics[height=5.4cm]{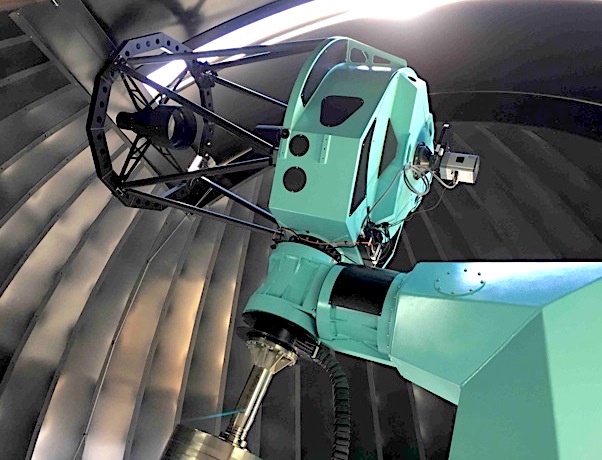}
\end{center}
\vspace{-0.2cm}
\caption{\textit{Left:} The four domes of the SPECULOOS Southern Observatory (credit: E. Jehin). \textit{Right:} Callisto, the third telescope of the facility, which is in the leftmost dome on the left picture (credit: D. Queloz).} 
\label{fig:sso}
\end{figure*} 

\section{The SPECULOOS Observatories}

\subsection{The SPECULOOS Southern Observatory}
\label{subsec:sso}

\begin{figure*} [ht]
\begin{center}
\vspace{-0.2cm}
\includegraphics[height=5.8cm]{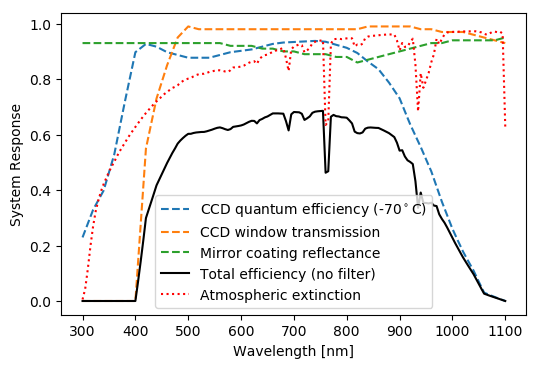}
\includegraphics[height=5.8cm]{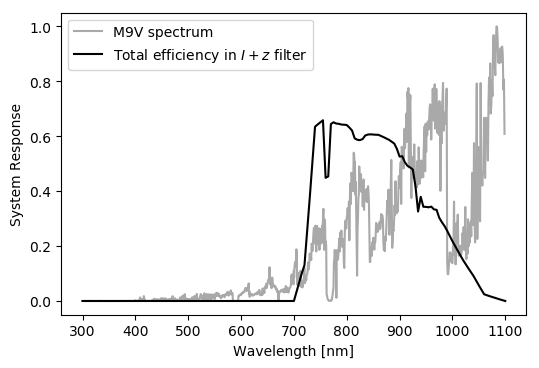}
\end{center}
\vspace{-0.4cm}
\caption{\textit{Left:} Overall system efficiency (black curve) taking into account the CCD quantum efficiency (blue dashed line), the CCD window transmission (orange dashed line), the combined reflectance curve of the primary+secondary mirrors (shown here for one mirror, green dashed line), and atmospheric extinction (for an airmass of 1.5, red dotted line), but assuming that the filter wheel is in ``clear'' position (no filter). \textit{Right:} Overall system efficiency through the $I+z$ filter (black curve) compared to a PHOENIX\cite{husser2013} synthetic M9-dwarf spectrum with $T_{\rm{eff}}$ = 2300 K (grey curve), scaled for clarity. The transmission curve of the $I+z$ filter is shown together with those of the other filters in Figure \ref{fig:filters}. The system efficiency beyond $\sim$800 nm is essentially set by the CCD quantum efficiency.}
\label{fig:response}
\end{figure*}

The SPECULOOS Southern Observatory is currently being commissioned at ESO Paranal Observatory (latitude: -24$^\circ$36'56.2'', longitude: -70$^\circ$23'25.4'', see Fig. \ref{fig:sketch} - top). Combining low humidity (80\% of nights with <4mm precipitable water vapor) and excellent photometric conditions (78\% of photometric nights)\footnote{\url{https://www.eso.org/sci/facilities/paranal/astroclimate/site.html}}, Paranal is an optimal site for our survey. SSO is composed of four robotic 1-m Ritchey-Chretien telescopes\footnote{Since we expect the typical planetary systems around UCDs to be scaled-up versions of the Jovian satellite system (with terrestrial planets replacing the Galilean moons), we decided to name the four telescopes Io, Europa, Ganymede, and Callisto.} built by the German ASTELCO\footnote{\url{http://www.astelco.com/}} company (see Fig. \ref{fig:sketch} - bottom). For each telescope, the 1-m diameter primary mirror has a F/2.3 focal ratio and is coupled with a 28-cm diameter secondary at a relative distance of $\sim$1.6 m, resulting in a combined F/8 focal ratio for the system. Both mirrors are covered with a raw aluminium coating, whose combined reflectance curve is shown in Figure \ref{fig:response} (left). The telescopes have a compact and open design with a lightweight optical tube assembly made of steel, aluminium and carbon fibre components (see Fig. \ref{fig:sso} - right). This design provides a good aerodynamic behavior and low wind resistance, allowing to observe at high wind speeds (up to $\sim$50km/h). Focusing is realized by a motorized axial movement of the secondary mirror, within a range of $\pm$20 mm with an accuracy of $\pm$5 $\mu$m. 

Each telescope is associated with a robotic equatorial New Technology Mount NTM-1000, also from ASTELCO. This mount uses direct drive torque motors, which allows very fast slewing (8$^\circ$/sec, up to 20$^\circ$/sec possible), accurate pointing (pointing accuracy better than 5'') and tracking (tracking accuracy without autoguider better than 1'' over 10 min), without periodic errors. Such a good tracking accuracy can be achieved thanks to the high encoder resolution (0.0029''/increment) that allows to minimize jitter effects. 

A key component of our strategy to achieve high photometric precisions is to keep the stars on the same few pixels for a whole exposure sequence. This is done using an updated version of the DONUTS autoguiding system described by McCormac et al. (2013)\cite{donuts}. Briefly, this technique relies on a reference image of each science field. The reference image is summed along the X and Y directions, creating two 1D reference image projections. A pair of 1D comparison projections are created for each subsequent science image and the guide correction is measured from a pair of cross-correlations between reference and comparison, in both X and Y directions. Initial tests on-sky have returned a guiding precision of $\leq 0.5$ pixels ($\leq 0.15$'') RMS over tens of nights. Further optimization may be possible to improve the current level of precision.

Each telescope is enclosed in a 6.25-m diameter circular building surmounted by an automatized hemispheric wide-slit dome (with sliding doors, see Fig. \ref{fig:sso} - left), made by the Italian Gambato\footnote{\url{http://gambato.it/it/}} company. The domes are made of aluminium and painted white outside to minimize reflections and internal heating during the day. A complete azimuth rotation of 360$^\circ$ takes less than 5 min. Each building also includes an additional rectangular shelter for storage of equipment and for hosting the telescope control computers. The dome and building sizes and distances from each other are optimized to ensure the non-vignetting of each other down to 20$^\circ$ above horizon.

Each telescope is equipped with an Andor\footnote{\url{http://www.andor.com/}} iKon-L thermoelectrically-cooled camera with a near-IR-optimized deeply depleted 2K$\times$2K e2v\footnote{\url{https://www.e2v.com/}} CCD detector (13.5 $\mu$m pixel size). The field of view on sky is 12'$\times$12', yielding a pixel scale of 0.35''/pixel. The camera can be cooled down to -100$^\circ$C (5-stage Peltier cooling). It is usually operated at -60$^\circ$C with a dark current of $\sim$0.1 e$^{-}$/s/pixel. The detector provides a good sensitivity from \hbox{$\sim$350 nm} (near-UV) to $\sim$950 nm (near-IR), with a maximum quantum efficiency of $\sim$94\% at both 420 and 740 nm (see Fig. \ref{fig:response} - left). However, the window of the camera is optimized for the visible/near-IR and blocks all wavelengths below $\sim$400 nm (see Fig. \ref{fig:response} - left). The camera also has a very low fringing in the near-IR ($<$1\%) thanks to both the wedge-design of the window and e2v proprietary fringe suppression technology applied to the detector. There are 4 readout speeds available, up to 5 MHz. SPECULOOS observations are usually performed using the 1MHz readout mode with a pre-amplifier gain of 2 e$^{-}$/ADU, providing a low readout noise of 6.2 e$^{-}$. The gain is 1.04 e$^{-}$/ADU with these settings.

\begin{figure} [ht]
\vspace{-0.4cm}
\begin{center}
\includegraphics[height=5.8cm]{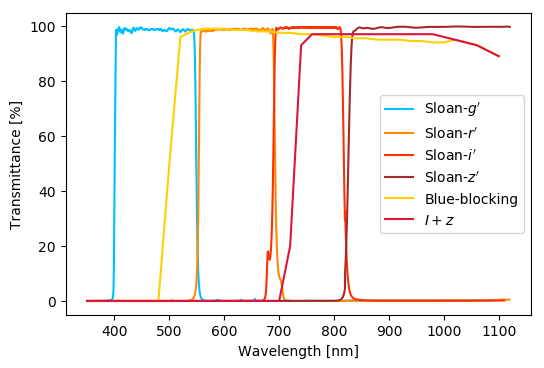}
\end{center}
\vspace{-0.4cm}
\caption{Transmission curves of the different broad-band filters currently available on the telescopes of the SPECULOOS Southern Observatory.}
\label{fig:filters}
\end{figure}

Each camera is associated with a filter wheel from Finger Lakes Instrumentation\footnote{\url{http://www.flicamera.com/}} (model CFW3-10) allowing 10 different 5$\times$5 cm filters. The following broad-band filters, all manufactured by Astrodon\footnote{\url{http://www.astrodon.com/}}, are currently available: Sloan-$g'$, -$r'$, -$i'$, -$z'$ filters, and special exoplanet filters ``$I+z$'' (transmittance >90\% from 750 nm to beyond 1000 nm) and ``blue-blocking'' (transmittance >90\% from 500 nm to beyond 1000 nm). The transmission curves of these filters are shown in Figure \ref{fig:filters}.

\setcounter{footnote}{0}

The observatory is fully robotic and can be controlled from anywhere where there is internet access, through a Virtual Private Network (VPN) connection between Paranal and the University of Li\`ege (Belgium). Observing plans, consisting of very simple text files, are submitted daily to the ACP Expert Observatory Control Software\footnote{\url{http://scheduler.dc3.com/}}\cite{acp} installed on the control computer of each telescope. ACP works in combination with the control softwares of the individual subsystems and handles all aspects of the observatory: control of the automated dome, pointing, filter wheel's management, image acquisition, focusing, guiding, etc. Each telescope is equipped with a Boltwood Cloud Sensor II weather station from the Diffraction Limited company\footnote{\url{http://diffractionlimited.com/}}, which monitors the cloud cover, wind speed, humidity, dew point, and amount of daylight. This weather station also includes a moisture sensor which detects rain and snow. It is connected to ACP, triggering an automatic closure of the dome in case of bad conditions. Each dome is also equipped with additional rain and light sensors, working independently from the telescope control computer for extra safety. Several IP-power sockets are connected to the electrical devices inside the domes, to allow remotely rebooting the systems if necessary. Finally, each telescope is also equipped with an uninterruptable power supply (UPS), as well as several webcams and microphones.

\subsection{The SPECULOOS Northern Observatory}
\label{subsec:sno}

\begin{figure*} [ht]
\begin{center}
\includegraphics[height=7.8cm]{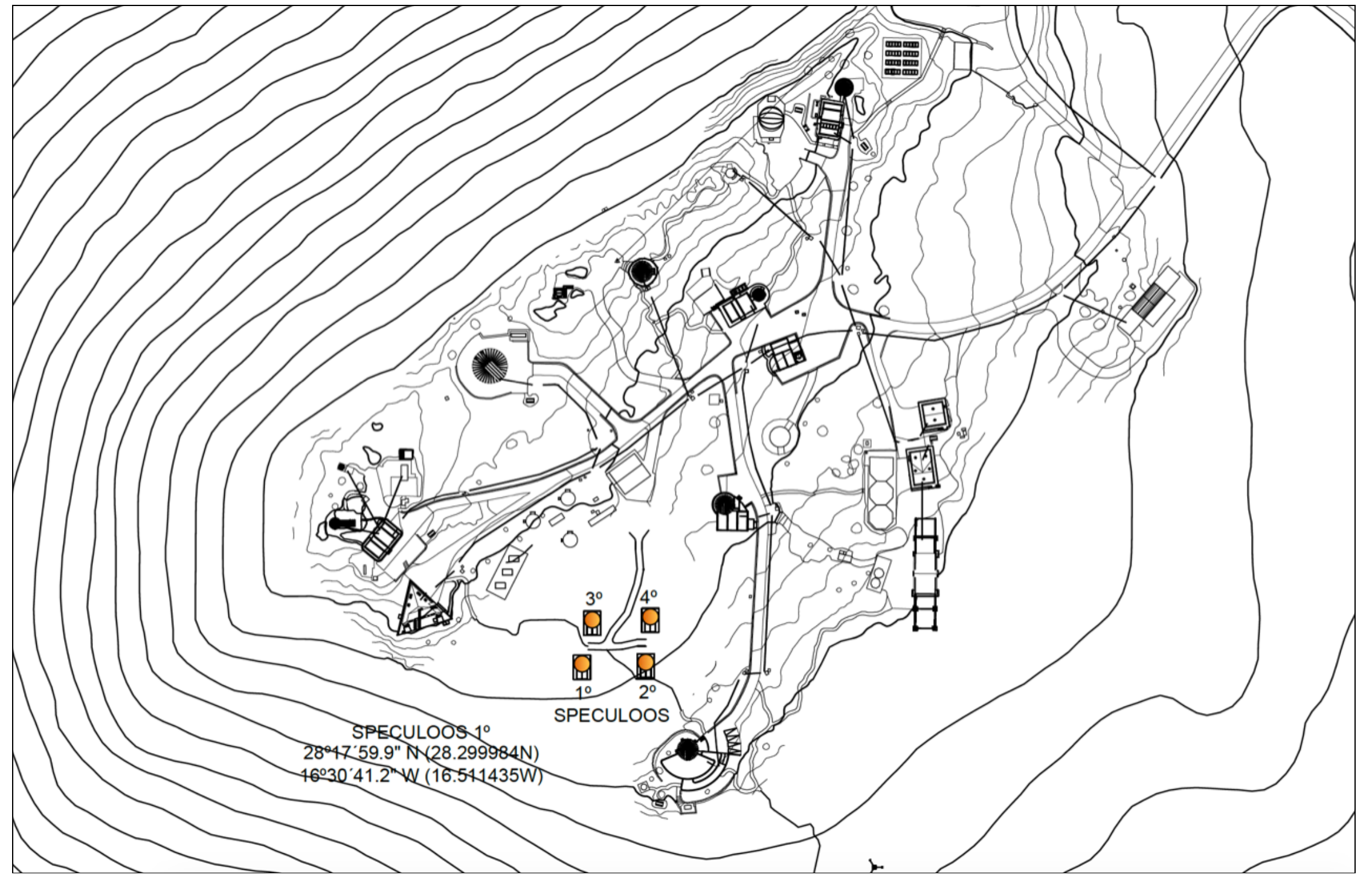}
\includegraphics[height=3.94cm]{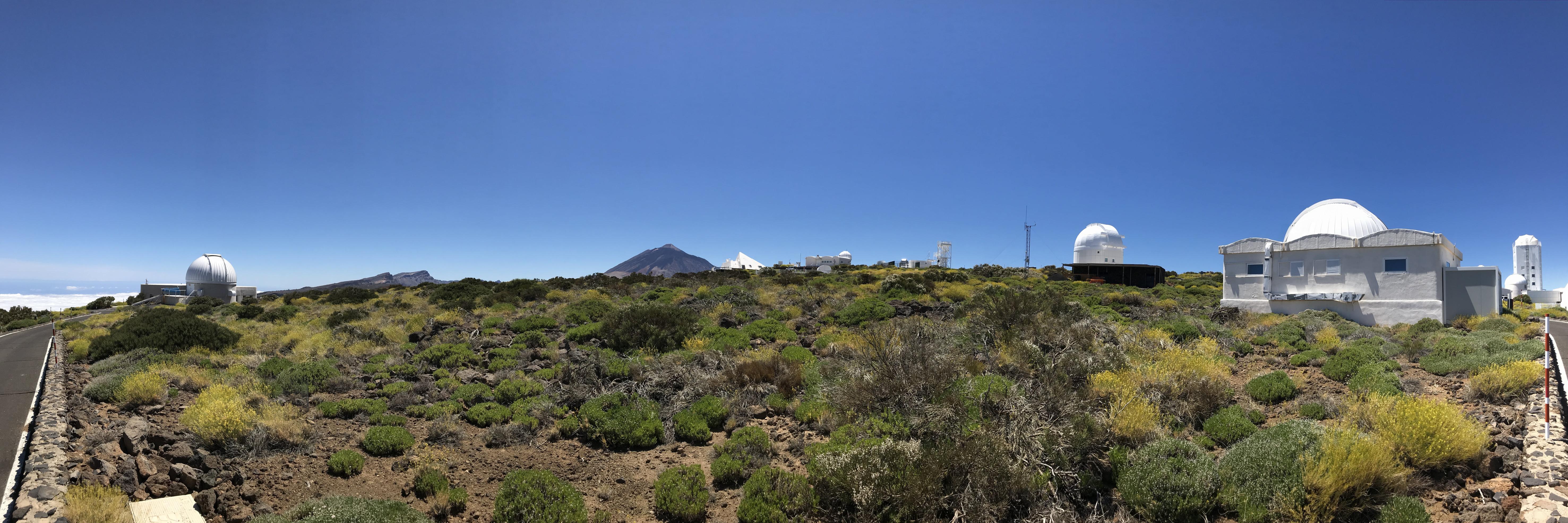}
\end{center}
\caption{\textit{Top:} Location of the SPECULOOS Northern Observatory at Teide Observatory (credit: Juan Jos\'e Saavedra Gallo). \textit{Bottom:} Panorama of the site reserved for SNO at Teide Observatory (Mount Teide visible at the back).}
\label{fig:sno}
\end{figure*}

The SPECULOOS Northern Observatory is currently being developed at Teide Observatory in Tenerife (latitude: 28$^\circ$17'59.9'', longitude: 16$^\circ$30'41.2'', see Fig. \ref{fig:sno}). SNO is envisioned as a twin observatory to SSO. As of June 2018, two of four telescopes are funded. The first of these telescopes, Artemis, will be commissioned over December 2018. Teide Observatory is a professional observatory managed by the Instituto de Astrof\'isica de Canarias (IAC) -- part of the University of La Laguna -- and located between the Teide National Park and the Corona Forestal Natural Park. It presents excellent weather conditions, combining low humidity and clear skies more than 250 nights per year. Its meteorological statistics are similar to its twin site of La Palma (previously the option B for the Thirty Meter Telescope). In addition, Teide Observatory presents unique qualities compared to the other excellent sites of the Northern Hemisphere investigated, notably at the financial, scientific, and operational levels for our teams located in the US and Europe, while allowing a quick deployment of the facility (first telescope to be operational within 18 months from first discussion with the observatory). The operation of SNO will mimic SSO's, leveraging its full robotization.

\section{Preliminary photometric performance}

\begin{figure} [ht]
\vspace{-0.3cm}
\begin{center}
\includegraphics[height=5.6cm]{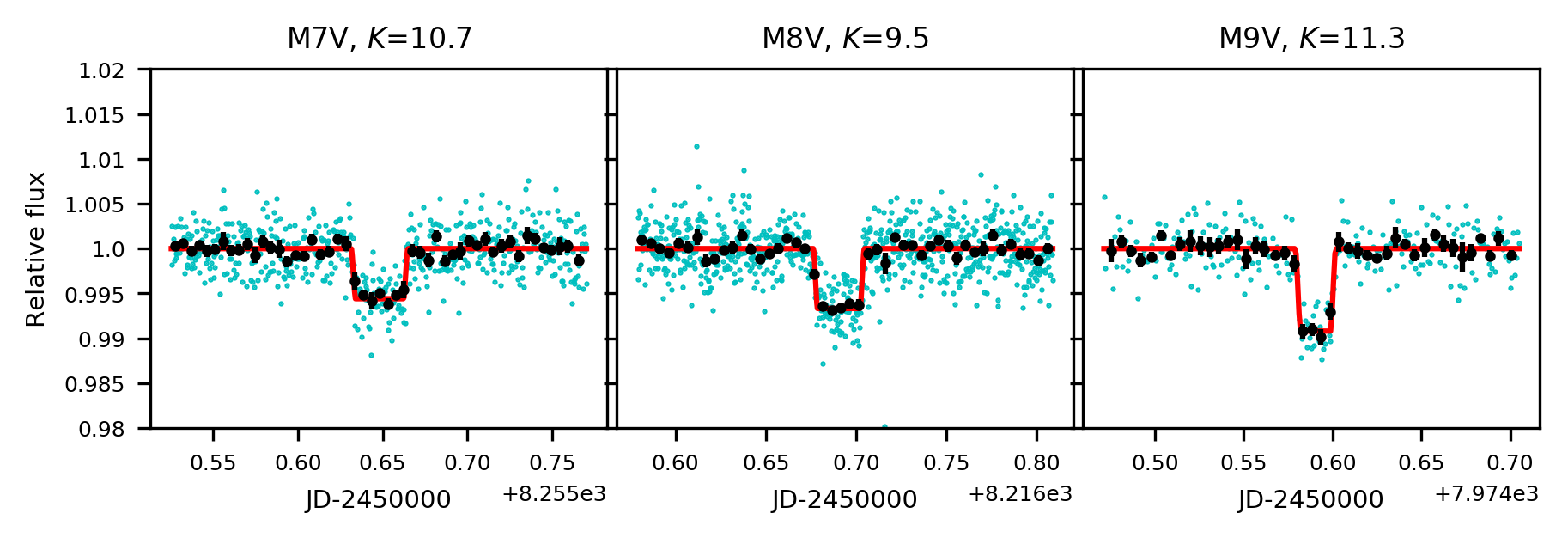}
\end{center}
\vspace{-0.5cm}
\caption{Light curves of three UCD stars obtained with the SPECULOOS Southern Observatory, after injection of a synthetic transit of a habitable Earth-size planet. For each light curve, the measurements are shown unbinned (cyan points) and binned per 7.2 min (black points with error bars), together with our best-fit transit model (red line).}
\label{fig:fake}
\end{figure}

\begin{figure} [!ht]
\vspace{-0.6cm}
\begin{center}
\includegraphics[height=9.9cm]{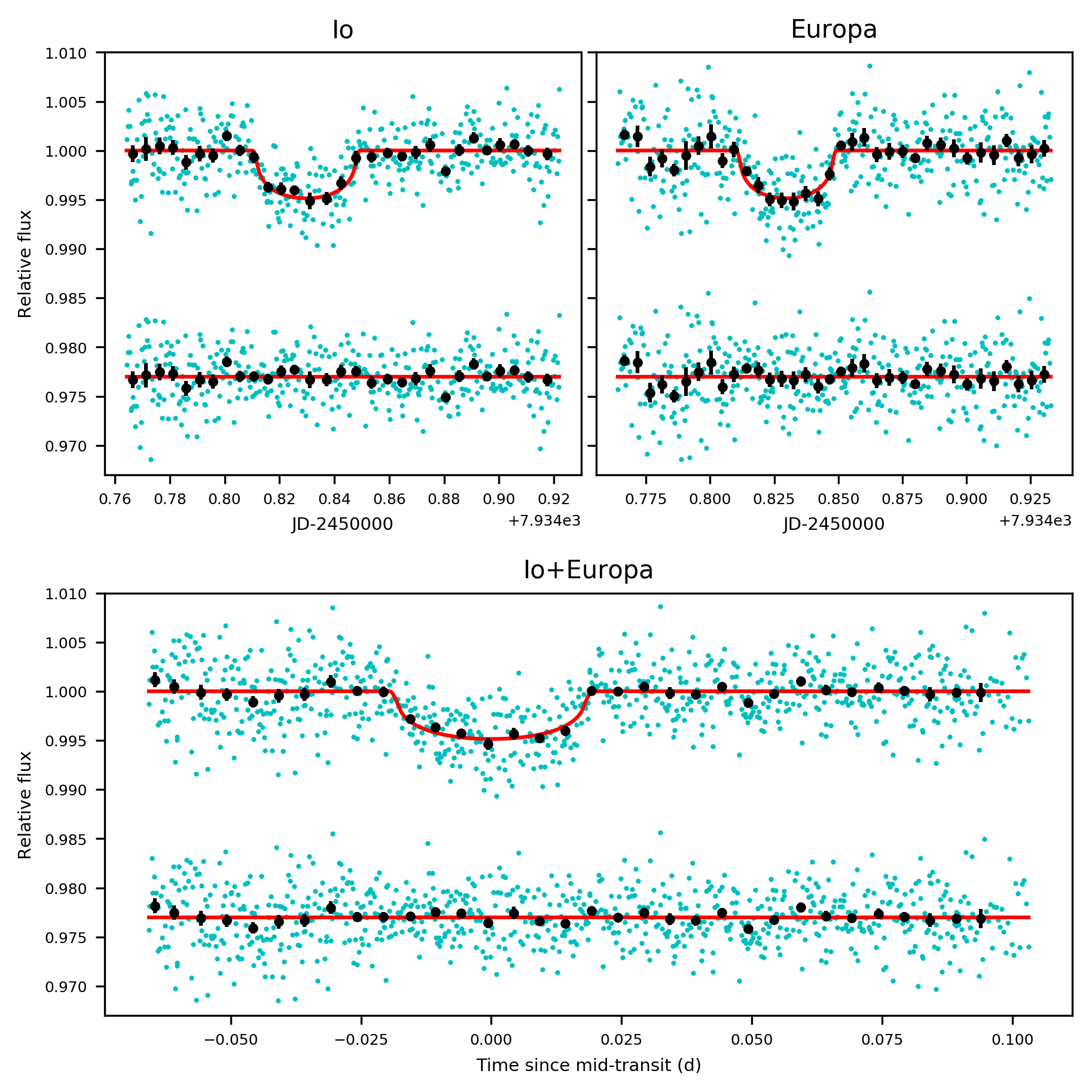}
\end{center}
\vspace{-0.4cm}
\caption{Transit of TRAPPIST-1e observed simultaneously with 2 SSO telescopes, Io and Europa, in June 2017. For both telescopes, the transit was observed in $I+z$ filter, with a 20s exposure time. The top two panels show the individual light curves, while the bottom panel shows the combined light curve. For each panel, the measurements are shown unbinned (cyan points) and binned per 7.2 min (black points with error bars), together with our best-fit transit model (red line). The best-fit residuals are also shown below each light curve, shifted along the $y$-axis for the sake of clarity.}
\label{fig:t1e}
\end{figure}

The last telescope of the SPECULOOS Southern Observatory will be installed at Paranal in summer 2018. Over the past few months, the first three telescopes have already started monitoring nearby UCDs, allowing us to assess the potential of the facility for the detection of transiting terrestrial planets around these objects. Figure \ref{fig:fake} shows some of our light curves for targets of different spectral types, after injection of a synthetic transit signal of a habitable Earth-size planet. It can be seen that the transits are easily visible to the naked eye. Analysing these transit light curves with the latest version of our adaptive MCMC code\cite{gillon2012}, we derived detection significances of 12, 14, and 17-$\sigma$ for the transits on the M7V, M8V, and M9V targets, respectively. Assuming a 5-$\sigma$ detection threshold, the detection limits in terms of planetary radius for these data are 0.68 $R_{\oplus}$ for the M7V, 0.65 $R_{\oplus}$  for the M8V, and 0.57 $R_{\oplus}$  for the M9V. As part of our ground-based follow-up campaign of the TRAPPIST-1 system (Ducrot et al. in prep.), we also observed transits of the TRAPPIST-1 planets with some SSO telescopes during the second half of 2017. Figure \ref{fig:t1e} shows a transit of TRAPPIST-1e that was observed simultaneously with 2 telescopes, Io and Europa, in June 2017. It can be seen that the $\sim$0.5\%-deep transit is readily detected in a single event. The detection significance is $\sim$10-$\sigma$ in the individual light curves and $\sim$14-$\sigma$ in the combined light curve. The RMS of the best-fit residuals is $\sim$770 ppm/7.2 min for the individual light curves and \hbox{$\sim$530 ppm/7.2 min} for the combined one. Together, these first results demonstrate the exquisite photometric precisions achievable with the SPECULOOS Southern Observatory on nearby UCDs and its capability to detect on these objects single transits of habitable planets of Earth-size and below, down to the Mars-size regime.

\begin{figure} [ht]
\begin{center}
\includegraphics[height=7cm]{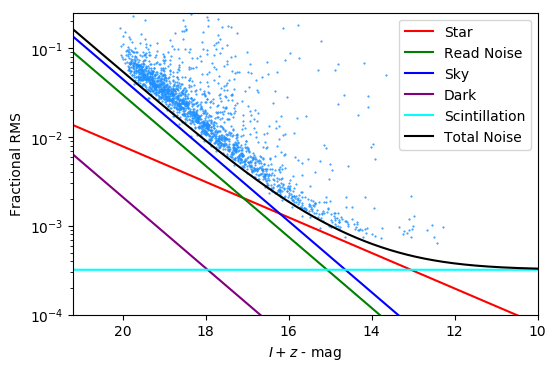}
\end{center}
\vspace{-0.3cm}
\caption{The data points show the measured fractional RMS of the differential light curves (binned per 10 min) of all the stars in a field observed during one night with one of the SSO telescopes, as a function of their apparent magnitude in the $I+z$ band. The black curve represents a noise model accounting for different contributions: Poisson noise from the star (red), read noise from the detector (green), noise from background light (blue), noise from dark current (purple), and atmospheric scintillation (cyan).}
\label{fig:resp_phot}
\end{figure}

Our commissioning data also allowed us to assess the global photometric performance of the SSO telescopes, i.e. not only for UCD targets but for any star, as a function of stellar brightness. Figure \ref{fig:resp_phot} shows the measured fractional RMS of the differential light curves (binned per 10 min) of all the stars in a field observed during one night with one of the SSO telescopes ($I+z$ filter, 10s exposure time), as a function of their apparent magnitude in the $I+z$ band. The data are compared with a noise model\cite{merline1995} accounting for different contributions: Poisson noise from the star, read noise from the detector, noise from background light, noise from dark current, and atmospheric scintillation. While we are still working on improving our reduction pipeline, that will be presented in a separate paper (Murray et al. in prep.), it can be noted that the data already match closely the noise model. For bright unsaturated stars, the precision is limited by atmospheric scintillation, for which we have adopted here the law for Paranal from Ref. \citenum{osborn2015}. Poisson noise from the star dominates for intermediate-brightness stars, while noise from background light dominates for fainter stars, with also a significant contribution from readout noise in that regime. The noise contribution from dark current is negligible. The photometric precisions (10-min binning) range from a few percents for the faintest detected objects, to a few mmag for intermediate-brightness stars, to $\sim$0.5 mmag for the brightest unsaturated objects.


\section{Expected yields}

\subsection{Estimated planet yield}
We performed Monte-Carlo simulations to estimate the detection potential of SPECULOOS. Each simulation looped within the current SPECULOOS target list (1136 UCDs with $K$-mag $\leq$ 12.5). For each target, a value for the mass, radius, and effective temperature was drawn from the normal distributions shown in Table \ref{table:simu}, and a typical observation window was drawn, assuming $N$ nights of observation in total - $N$ depending on the spectral type (see Section \ref{sec:concept}) - and a continuous observation sequence of 8 hours per night. A planetary system was then drawn for the star, based on the structure of the TRAPPIST-1 system. The orbital periods were divided in 8 possible bins: 1-2 d, 2-3.5 d, 3.5-5.5 d, 5.5-8 d, 8-11 d, 11-14.5 d, 14.5-18.5 d, and 18.5-23 d. For each period range, the probability for a planet to exist was fixed to 50\%, and 80\% if a planet was already present in a shorter orbit, assuming thus that low-mass stars tend to form compact multiple systems of low-mass planets\cite{ormel2017} and that each of our targets has at least one planet in a short-period orbit. An orbital period was attributed to each planet assuming a uniform probability distribution within its given period range. The orbits were assumed to be circular, and semi-major axes were computed using Kepler's third law. The mass and radius of each planet were drawn from the normal distributions shown in Table \ref{table:simu}, based on the mean and standard deviations of the masses and radii of the seven TRAPPIST-1 planets \cite{grimm2018}. A  value for the cosine of the orbital inclination $\cos{i}_1$ of the inner planet of the system was drawn from the uniform distribution $U(0,1)$. Then inclinations were drawn for the outer planets from the normal distribution $N(i_1,0.2^2)$, assuming thus a nearly coplanar system.

For each planet, the impact parameter at inferior conjunction was computed. If smaller than 1, the depth and duration of the transit were computed, a random phase was drawn for the orbit, and the code  checked if a transit happened within the observation window. If a transit was observed, a photometric signal-to-noise was computed based on its duration and on the expected photometric precision of a SPECULOOS 1m telescope for the star, given its spectral type and magnitude, as computed using the SPECULOOS-South Observatory Exposure Time Calculator. A noise floor level of 500 ppm was assumed to represent the red noises of instrumental, atmospheric, and astrophysical origins. At the end, if the resulting signal-to-noise was larger than 5, the transit was assumed to be detected, and the parameters of the planet and the star registered. 

\begin{figure*} [ht]
\begin{center}
\includegraphics[height=16cm]{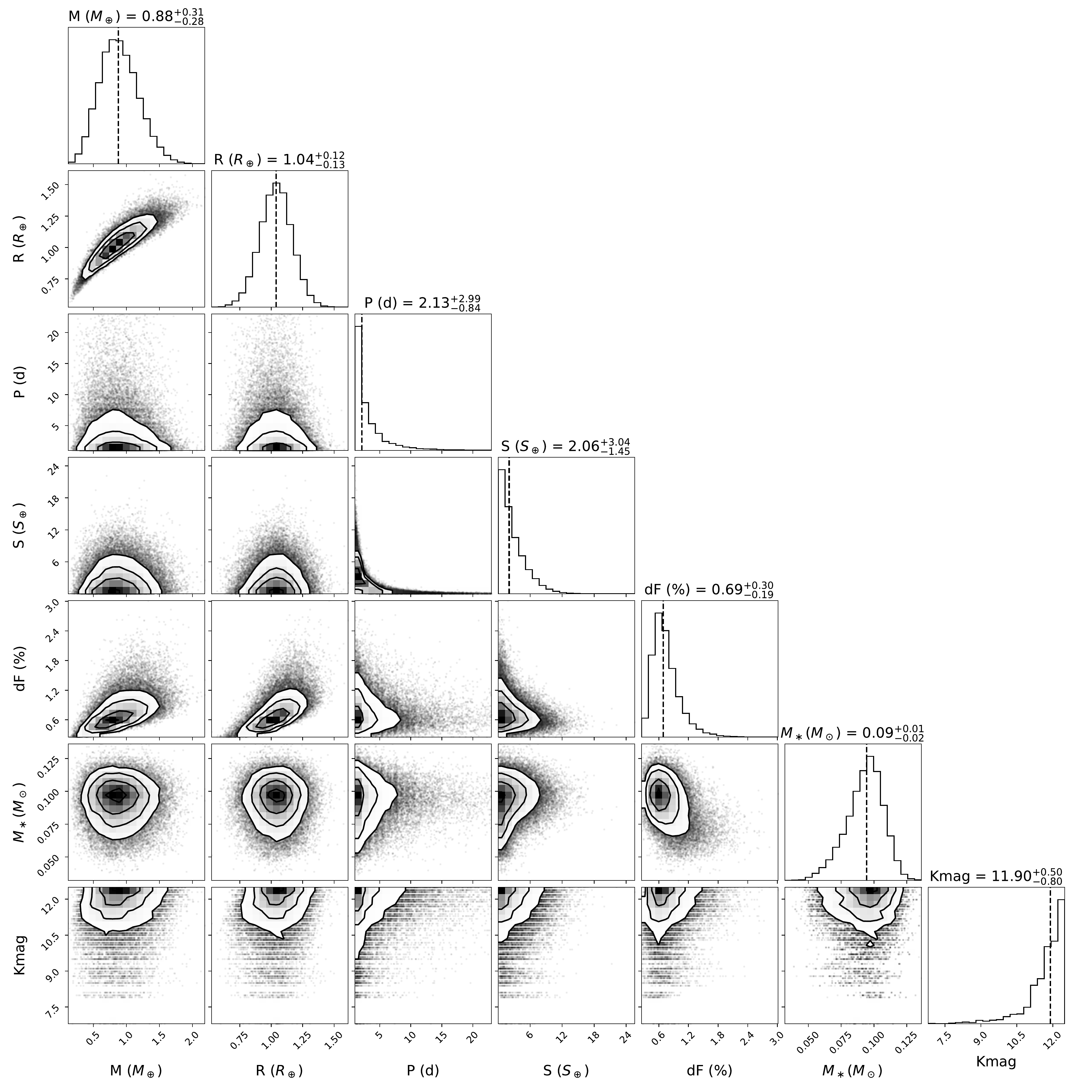}
\end{center}
\vspace{-0.2cm}
\caption{Corner plot showing the 1D and 2D histograms for the parameters of the detected planets and their host stars as predicted by our yield simulations (Section 7.1). The 1-, 2-, and 3-$\sigma$ contours are shown for each 2D histogram. $M$, $R$, $P$, and $S$ are the planetary mass, radius, orbital period, and irradiation, respectively. $dF$ is the transit depth. $M_\ast$ is the host star's mass, and Kmag is its magnitude in the $K$-band. Figure done with \texttt{corner.py}\cite{corner}.}
\label{fig:simu_corner}
\end{figure*}

We performed 1000 simulations that resulted for the detected planets in the probability distributions shown in Figure \ref{fig:simu_corner}. Planets with irradiations between 25 and 150\% times the one of the Earth were assumed to be temperate\cite{kopparapu2013}. Table \ref{table:simu} summarizes the results of the simulations. Under our work assumptions, SPECULOOS should detect $\sim$40 planets - including $\sim$15 temperate ones - in $\sim$20 systems. With 8 telescopes and a global efficiency of 70\%, the survey should take $\sim$10 years to be completed. 

Figure \ref{fig:simu_corner}  shows that SPECULOOS will be well sensitive to Earth-sized planets, its most likely detected planet being an Earth-sized world on a $\sim$2d orbit around a $\sim$0.09 $M_\odot$ star with a $K$-mag around 12. This harvest has of course to be taken with a pinch of salt, as the planetary population of UCDs is still poorly explored. Nevertheless, there are only 81 objects with $K$-mag $\leq$ 10.5 in the SPECULOOS target list, including TRAPPIST-1 ($K$-mag = 10.3). Running the same simulations than described above for this brightest sub-sample leads to $\sim$1-6 detected planets - including 0-2 temperate ones - within $\sim$1-3 systems. The detection of TRAPPIST-1 is thus consistent with a high frequency (a few dozens of \%) of compact systems of Earth-sized planets around UCD stars. But we note that even when assuming a frequency of 100\% for planets in each period ranges, only 4.5\% of the simulations lead to the detection of a system of 7 planets transiting a target brighter than $K$-mag = 10.5. Nature was thus very kind to us, Earth people, by adjusting the ecliptic plane of the TRAPPIST-1 system to make it so well edge-on as seen from our planet. 

\begin{table}[ht]
\centering                          
\begin{tabular}{| c c c c |}        
\hline               
Spectral type  & Mass ($M_\odot$)& Radius ($R_\odot$) & $T_{\mathrm{eff}}$ (K) \\    
\hline                        
 M7-M7.5 &  $N(0.1,0.01^2)$ & $N(0.12,0.01^2)$ & $N(2700,200^2)$ \\     
 M8-M8.5  & $N(0.09,0.01^2)$  & $N(0.11,0.01^2)$ & $N(2500,200^2)$ \\  
 M9-M9.5 & $N(0.08,0.01^2)$  & $N(0.1,0.01^2)$ & $N(2300,200^2)$ \\  
 L &  $N(0.07,0.01^2)$ &  $N(0.09,0.01^2)$ & $N(2100,200^2)$ \\  
\hline
\hline
& Planet mass ($M_\oplus$) & $N(0.81,0.34^2)$  & \\
& Planet density ($\rho_\oplus$) & $N(0.79,0.12^2)$  &  \\ 
\hline\hline
& Planets & $42 \pm 10$ & \\
& Systems & $22 \pm 5 $  & \\
& Temperate planets & $14 \pm 5$ & \\
\hline                               
\end{tabular}
\vspace{0.1cm}
\caption{Normal distributions of the stellar (\textit{top}) and planetary (\textit{middle}) parameters assumed in our SPECULOOS yield simulations, and resulting harvest (\textit{bottom}).}             
\label{table:simu}
\end{table}

\subsection{Extra scientific outputs}

\begin{figure*} [ht]
\vspace{-0.1cm}
\begin{center}
\includegraphics[height=5.94cm]{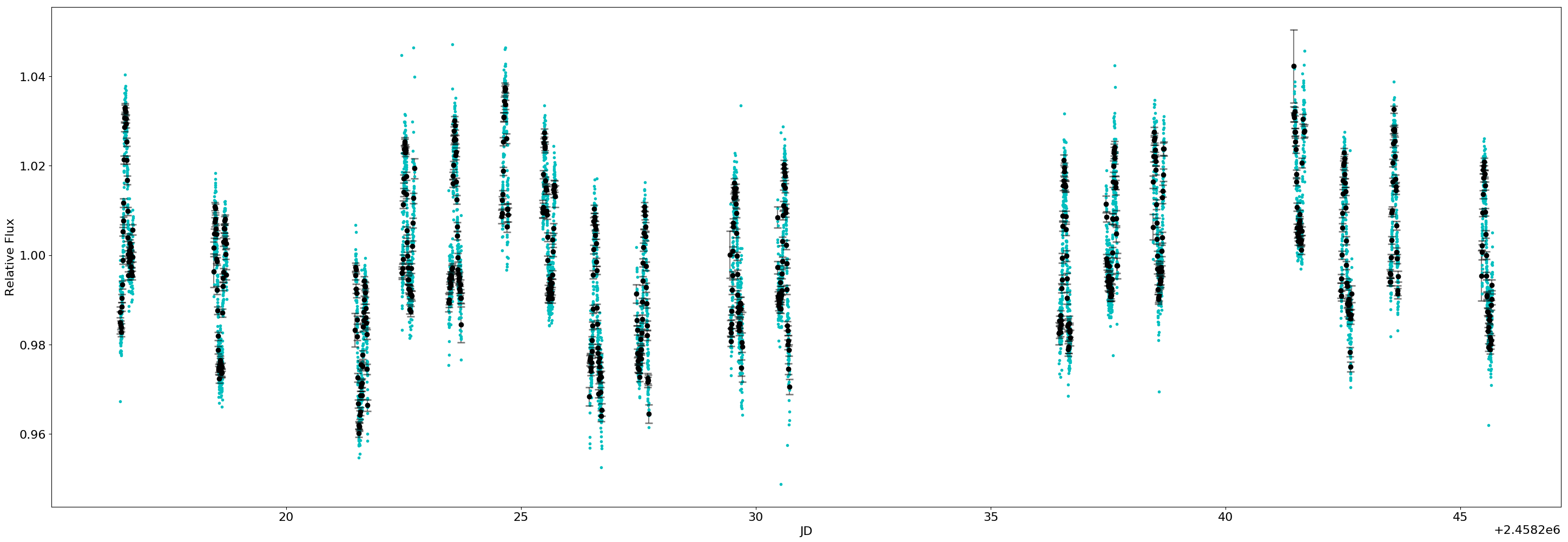}
\includegraphics[height=6.5cm]{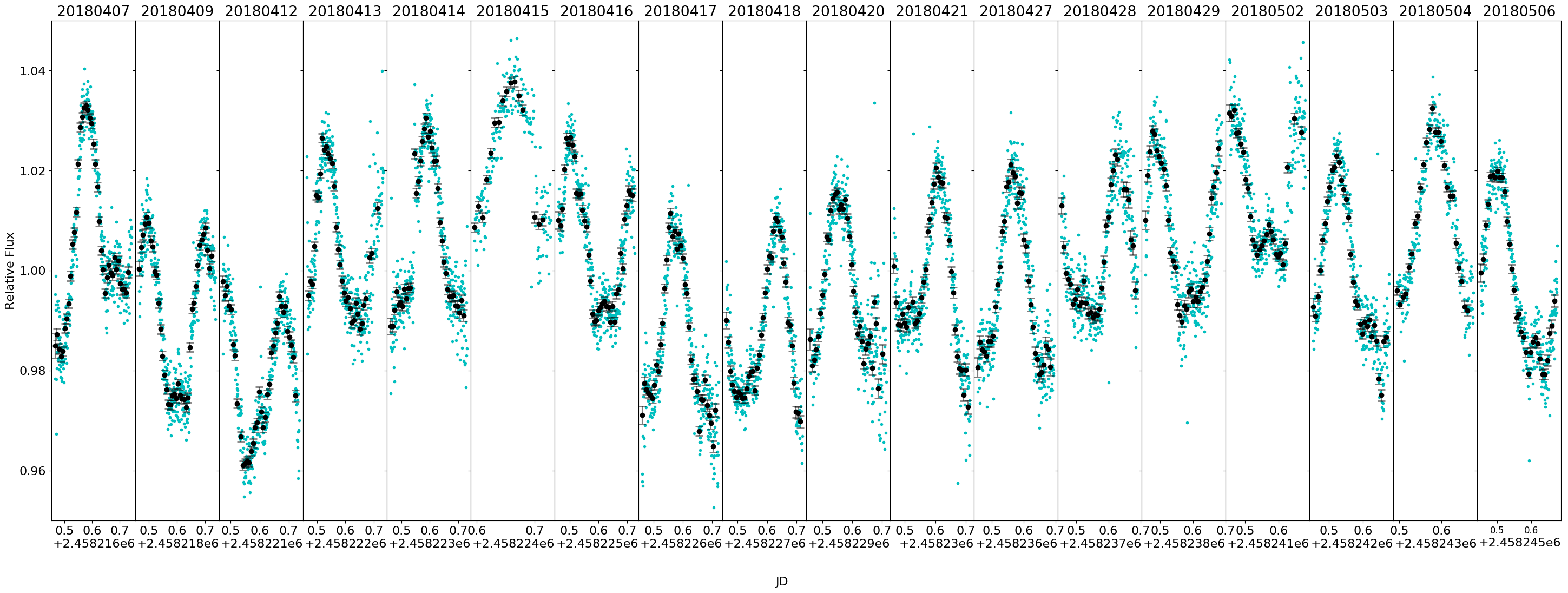}
\end{center}
\vspace{-0.3cm}
\caption{Global differential light curve of one of our targets, an M8-type UCD, observed with one telescope of the SPECULOOS-Southern Observatory (Callisto) over 18 nights in April-May 2018. This target shows some evolving multi-period variability with a maximal amplitude of $\sim$8\%. The top panel shows the global differential light curve over 18 nights, while the lower panel shows individual zooms on each night. The measurements are shown unbinned (cyan points) and binned per 10 min (black points with error bars).}
\label{fig:sp1003}
\end{figure*}

By gathering long sequences of high-precision near-IR photometry for nearby UCDs, SPECULOOS also represents a unique tool for exploring the physics of these objects. UCDs are generally rapidly rotating bodies with typical periods of only a few days for UCD stars and periods <24 hours for most brown dwarfs\cite{blake2010,metchev2015}. As mentioned in Section \ref{sec:prototype}, UCDs can show some rotational photometric variability (see Fig. \ref{fig:ucdts} - third target), induced by spots or heterogeneous cloud decks for the cooler objects. In this context, the SPECULOOS photometric monitoring of these objects over a duration of at least 10 nights, is particularly valuable as it covers several full rotations (see Figure \ref{fig:sp1003} for an example). For the UCD stars of our sample, it makes possible the precise determination of their rotation periods, the estimation of their fractional surface coverage of active regions, and the exploration of the possible relationship between activity and rotation period\cite{reid2005,williams2013}. In the brown-dwarf regime, the long-duration photometric monitoring provided by SPECULOOS can help disentangle their rotation periods from complex effects due, e.g., to differential rotation, atmospheric circulation, or cloud evolution, ultimately leading to new insights into their atmospheric physics\cite{gillon2013luhman,street2015,apai2017}. Finally, coupling the flares frequency of UCDs (see Fig. \ref{fig:ucdts} - second target) in SPECULOOS photometry to archival UV, X-ray, and radio measurements can provide an estimate of the magnetic energy output from these objects\cite{berger2005}.

In addition to producing a unique photometric database for a large sample of UCDs, the SPECULOOS data will also represent a valuable dataset for the astronomical community at large: the intense photometric monitoring ($\geq$10 nights) with 1-m telescopes of a total field of view\footnote{Total field of view obtained by combining our 1200 target fields, each of 12'$\times$12'.} of 48 deg$^2$ in the very-near-IR, with excellent time and spatial sampling. In agreement with ESO, we will provide a public access to the photometric data gathered with the SPECULOOS Southern Observatory after a 1-year proprietary period. Reduced images, as well as extracted light curves of all objects with \textit{J}-magnitude$\leq$15, will be made available to the community via the ESO archive\footnote{\url{http://archive.eso.org/cms.html}}. 

\section{Project organization}

SPECULOOS is a project led by the University of Li\`ege (Belgium) and carried out in collaboration with MIT (USA), the Universities of Cambridge (UK), Jeddah (Saudi Arabia), Bern (Switzerland), Birmingham (UK), California San Diego (USA), Warwick (UK), Cadi Ayyad (Morocco), and the Canary Islands Institute of Astrophysics (Spain). It is primarily funded by the European Research Council, the Simons and Heising-Simons Foundations, as well as several private sponsors.

\acknowledgments 
 
The research leading to these results has received funding from the European Research Council (ERC) under the FP/2007-2013 ERC grant agreement no. 336480, and under the H2020 ERC grant agreement no. 679030; and from an Actions de Recherche Concert\'ee (ARC) grant, financed by the Wallonia-Brussels Federation. This work was also partially supported by a grant from the Simons Foundation (PI Queloz, grant number 327127), as well as by the MERAC foundation (PI Triaud). TRAPPIST-South is a project funded by the Belgian Fonds (National) de la Recherche Scientifique (F.R.S.-FNRS) under grant FRFC 2.5.594.09.F, with the participation of the Swiss National Science Foundation (FNS/SNSF). TRAPPIST-North is a project funded by the University of Li\`ege, and performed in collaboration with Cadi Ayyad University of Marrakesh. LD acknowledges support from the Gruber Foundation Fellowship. VVG and MG are F.R.S.-FNRS Research Associates. JdW is grateful for the financial support received for the SPECULOOS Project from the Heising-Simons Foundation, P. Gilman, and C. \& L. Masson. EJ is F.R.S.-FNRS Senior Research Associate. B-OD acknowledges support from the Swiss National Science Foundation in the form of a SNSF Professorship (PP00P2\_163967). AJB acknowledges funding support from the US-UK Fulbright Scholarship programme.

\bibliography{biblio} 
\bibliographystyle{spiebib} 

\end{document}